\input  phyzzx
\input epsf
\overfullrule=0pt
\hsize=6.5truein
\vsize=9.0truein
\voffset=-0.1truein
\hoffset=-0.1truein

%
%

\def\IC{{\ \hbox{{\rm I}\kern-.6em\hbox{\bf C}}}}
\def\IR{{\hbox{{\rm I}\kern-.2em\hbox{\rm R}}}}
\def\IZ{{\hbox{{\rm Z}\kern-.4em\hbox{\rm Z}}}}

\def\sIR{{\hbox{{\sevenrm I}\kern-.2em\hbox{\sevenrm R}}}}

\def\Sc{Schwarzschild}

%
%
\hyphenation{Min-kow-ski}

\rightline{SU-ITP-97-38}
\rightline{September  1997}
\rightline{hep-th/9709225}

\vfill

%
%
\title{\bf Six Dimensional Schwarzschild Black Holes in M(atrix) Theory }

\bigskip

%
%

\author{Edi Halyo\foot{e-mail: halyo@dormouse.stanford.edu}}

\medskip

\address{Department of Physics \break Stanford University \break
 Stanford, CA 94305-4060}

\vfill

%
%
We calculate the entropy of six dimensional \Sc \ black holes in matrix theory.
We use the description of the matrix model on $T^5$ as the world--volume theory
of
NS five--branes and show that the black hole entropy is reproduced by
noncritical
closed strings with fractional tension living on the five--brane.

\vfill
\endpage
%
%
\REF\LE{L. Susskind, hep-th/9309145.}
\REF\HRS{E. Halyo, A. Rajaraman and L. Susskind, hep-th/9605112.}
\REF\HKRS{E. Halyo, B. Kol, A. Rajaraman and L. Susskind, hep-th/9609075.}
\REF\MM{T. Banks, W. Fischler, S. Shenker and L. Susskind, hep-th/9610043.}
\REF\DVV{R. Dijkgraaf, E. Verlinde and H. Verlinde, hep-th/9704018.}
\REF\LM{M. Li and E. Martinec, hep-th/9703211; hep-th/9704134.}
\REF\EDI{E. Halyo, hep-th/9705107.}
\REF\BFKS{T. Banks, W. Fischler, I. Klebanov and L. Susskind, hep-th/9709 }
\REF\KS{I. Klebanov and L. Susskind, hep-th/9709 }
\REF\DLCQ{L. Susskind, hep-th/9704080.}
\REF\NATI{N. Seiberg, hep-th/9705221.}
\REF\MALD{J. Maldacena, hep-th/9605016.}
\REF\DV{R. Dijkgraaf, E. Verlinde and H. Verlinde, Nucl. Phys. {\bf B486}
(1997) 77, hep-th/9603126; Nucl. Phys. {\bf B486} (1997) 89, hep-th/9604055.}
\REF\HAL{E. Halyo, hep-th/9610068; hep--th/9611175.}
\REF\KT{I. Klebanov and A. Tseytlin, Nucl. Phys. {\bf B475} (1996) 165,
hep-th/9604089.}
\REF\FHRS{W. Fischler, E. Halyo, A. Rajaraman and L. Susskind, hep-th/9703102.}


%
%

In recent years there has been great progress in understanding the microscopic
structure of
extreme and near--extreme black holes; however the same cannot be
said for \Sc \ black holes. The reason for this is the fact that \Sc \ black
holes are not
even nearly supersymmetric systems. Thus the powerful methods of supersymmetry
and D brane physics are not applicable in this case. Nevertheless, there has
been some understanding of  \Sc \ black holes in string theory[\LE,\HRS,\HKRS].
In ref. [\HRS]
it was shown that the black hole entropy in any dimension can be obtained if
one assumes
that they are described by highly excited fundamental strings and takes the
gravitational redshift of the string mass into account.
The microscopic description of
black holes requires knowledge of strongly coupled
string theory since these are not perturbative systems. At the moment the only
candidate for
nonperturbative string theory (M theory) is the matrix model[\MM]. Thus it is
important to
understand black holes in the matrix model and see if this enhances our
understanding of the
problem beyond string theory. Extreme and near--extreme matrix black holes in
five dimensions
have already been considered in [\DVV,\LM,\EDI]. \Sc \ black holes in matrix
theory in eight and other dimensions also have been recently investigated in
[\BFKS,\KS].

In this letter, we consider six dimensional \Sc \ black holes in the matrix
model in the Discrete Light--Cone Quantization (DLCQ) formulation[\DLCQ]. Since
the matrix model is formulated in a highly boosted frame with a large
light--cone momentum, the original \Sc \ black hole becomes a near--extreme
system. We describe
the matrix model compactified on $T^5$ as the world--volume theory of $N$  NS
five--branes
(of IIA string theory) with a transverese compact circle[\NATI]. The five brane
world--volume theory contains strings
whose tension is fixed by the size of the transverse circle (since these are
membranes which wrap around the circle and end on the five--brane). The D=6 \Sc
\ black hole is now
described by the deviation from extremality of  $N$ NS five--branes. The
original
longitudinal momentum modes (partons) in space--time become NS five--branes and
the \Sc \ black hole is described by non--BPS excitations of this system.

On the other hand, it is known that the entropy
of a near--extreme five--branes is given by a noncritical, non--BPS closed
string which lives on the $5+1$ dimensional world-volume[\DV,\MALD]. In order
to get the proper thermodynamic description
of the black hole one assumes that the tension of the string is fractional. In
the following we will assume that the strings which
give the five--brane entropy are the same ones that appear as membranes which
end on the NS five--branes. We will see that this reproduces the correct black
hole entropy up to numerical factors and shows how
the string tension becomes fractional. One problem with this approach is the
fact that the strings
must be weakly interacting for entropy to workout correctly. However, it is
known that the
noncritical strings living on the five--branes are self--dual and therefore
strongly coupled.

We consider the matrix model in the DLCQ formulation,
i.e. for finite light--cone momentum $P=N/R$ where $N$ and the light--cone
radius $R$
are finite. The lowest possible value of $N$ needed to count the black hole
entropy
is given by an argument from ref. [\BFKS]. Consider a black hole with radius
$R_s$ in its rest frame which is also extended along the light--cone direction.
Now it is possible that $R_s>R$ and the black hole will not fit in the
light--cone direction.
However, one can boost the black hole along the longitudinal direction and then
its longitudinal size contracts to
$$R^{\prime} \sim {M \over P} R_s={MR \over N} R_s  \eqno(1) $$
We see that in order for the black hole to fit into longitudinal size $R$ the
boost must be at least
$$N>M R_s \eqno(2)$$

For a \Sc \ black hole in $D$ space--time dimensions the radius is given by
($G_D=\ell_{11}^9/L^{11-D}$ is the $D$ dimensional Newton constant.)
$$R_s \sim (G_D M)^{1/{D-3}}  \eqno(3) $$
so that the boost parameter becomes
$$N \sim G_D^{1/{D-3}} M^{{D-2}/{D-3}}  \eqno(4) $$
which is the \Sc \ black hole entropy in $D$ dimensions! In the following we
will use this
value of $N$ in our calculations. One can interpret this as the minimal value
of $N$ that can result in the resolution
needed to count the black hole entropy. Note that with this amount of boost the
original \Sc
\ black hole (which was a highly nonextreme system) becomes near--extreme. The
system now
looks like $N$ partons of the matrix model which are slightly excited.

We consider $D=6$ \Sc \ black holes which means that we compactify the matrix
model on $T^5$.
For simplicity we compactify on $T^5$ with all sides equal to $L$.
This compactification of the matrix model is described by the world--volume
theory of $N$
coincident  NS five--branes (of IIA string theory) with a compact transverse
direction. The size of the five--brane world--volume
is given by
$$\Sigma_{1,2,3,4}={\ell_{11}^3 \over {RL}} \qquad \Sigma_5={\ell_{11}^6 \over
{RL^4}}
\eqno(5) $$
and the radius of the transverse circle is
$$M_s^2={{R^2 L^5} \over \ell_{11}^9} \eqno(6) $$
The $5+1$ dimensional world--volume theory has (2,0) supersymmetry. The sixteen
BPS states of the matrix model on $T^5$
can be realized as zero, two and four branes inside the NS five branes. In
addition, the
longitudinal membranes and five--branes are given by momentum modes and wound
BPS strings on the five--brane. All these states are bound to the five--branes
since the IIA string coupling is vanishing. The moduli space is given by
$${\cal M}={(R^4 \times S^1)^N \over S^N} \eqno(7) $$
and the system is at the singular point of the moduli space at which all the
five--branes
coincide.

The \Sc
\ black hole is a non--BPS excitation of the $N$ NS five--branes. It is known
that the entropy of
near--extreme five--branes is given by that of noncritical closed strings
living on
the five--brane world--volume. These ``instantonic strings" satisfy
$${c_{eff} \over 6} {T \over T_{eff}}=N \eqno(8) $$
In order to get the correct thermodynamics for low energy states of the black
hole one
assumes that the ``instantonic strings fractionate", i.e $c_{eff}=6$ and
$T_{eff}=T/N$.
The Hagedorn temperature of these strings with fractional tension gives the
Hawking temperature of the black hole.
In the $T^5$ compactification of the matrix model there are also strings
which are membranes wrapped around the transverse circle ending on the
five--brane. Their tension is $M_s^2$. These are the strings which will give
the black hole entropy. The string tension
fractionates if we assume that in the black hole regime the NS five--branes are
equally separated from each other rather than being on top of each other. This
configuration corresponds to a
nonsingular point in the moduli space.
Then, the distance between consecutive
five--branes becomes $M_s^2/N$ and therefore the tension of the strings become
fractional.
This kind of tension fractionation also plays a role in extreme and
near--extreme
black holes[\EDI].

We can now calculate the entropy of the $D=6$ \Sc \ black hole.
{}From eq. (4) we find that
$$N \sim M^{4/3} G_6^{1/3} \sim S \eqno(9) $$
For the noncritical strings living in six dimensions
$$E^2={n \over N} M_s^2 \eqno(10) $$
where $n$ is the oscillator level and the factor $N$ appears due to
fractionation of the string tension.
The entropy of the string is
$$S = 2\pi \sqrt { {c \over 6} n} \eqno(11) $$
The world--volume energy of the string corresponds to the light--cone energy in
space--time which means
$$E={R \over N} M^2 \eqno(12) $$
Using eqs. (9), (10) and (12) the entropy becomes
$$S \sim N G_6^{1/2} R M_s^{-1}  \eqno(13) $$
Noting that the six dimensional Newton constant is $G_6=L^5/\ell_{11}^9$ and
using eq. (6) for
$M_s$ we find
$$S \sim N \sim M^{4/3} G_6^{1/3} \eqno(14) $$
which is the correct six dimensional black hole entropy up to numerical
factors. We see that the black hole entropy is reproduced correctly by a string
living on the five--branes. This string is the boundary
of a membrane stretched between two five--branes a distance $M_s^2/N$ apart and
therefore
has fractional tension. In order to use the string entropy formula, eq. (11) we
assumed that these strings are weakly interacting. However, since these strings
are self--dual their coupling is
necessarily one. We do not know an explanation for this fact.

The six demensional black hole was also considered in ref. [\KS] where the
entropy was
obtained by taking $N$ coincident near--extreme D five--branes. Since the
world--volume theory is now a $5+1$ dimensional $U(N)$ SYM theory[\FHRS] the
entropy will be given by a gas of particles
and not strings[\KT].
Then the equation of state for a gas of particles in six dimensions is given by
$$S \sim \sqrt{N} E g_6 \eqno(15) $$
where $E$ is given by eq. (12) and the gauge coupling constant is
$$g_6^2={\ell_{11}^9 \over {R^2 L^5}}=M_s^{-1} \eqno(16) $$
Using the above and eq. (9) we find that this equation of state is the same as
that in eq. (13) arising from strings. The two descriptions of the six
dimensional black hole are related by U duality. In fact in addition to the
decsription of the matrix model on $T^5$ by NS five--branes of IIA string
theory there is another T dual description interms of NS five--branes of IIB
string theory
(which are S dual to D five--branes). The world--volume theory in this case is
a  $U(N)$ SYM theory with $(1,1)$ supersymmetry. However, this description is
only correct at low world--volume energies. It still reproduces the correct
black hole entropy since entropy is
an ultraviolet effect in space--time and therefore it is related to infrared
physics on the
brane world--volume.

\bigskip
\centerline{Acknowledgements}
We would like to thank Tom Banks, Eva Silverstein and Lenny Susskind for
discussions.

\vfill

\refout

\end
\bye